\documentclass[%
reprint,
amsmath,amssymb,
aps,
pra,
]{revtex4-2}
\usepackage[colorlinks=true, allcolors=blue]{hyperref}
\usepackage{graphicx}
\usepackage{dcolumn}
\usepackage{bm}
\usepackage{dsfont}
\usepackage{hologo}
\usepackage{physics}
\usepackage{lipsum}
\usepackage{url}
\usepackage{mathrsfs}

\newcommand{\N}{\mathds{N}}
\newcommand{\Z}{\mathds{Z}}

\newcommand{\R}{\mathds{R}}

\newcommand{\C}{\mathds{C}}

\renewcommand\abs[1]{\left\vert#1\right\vert}

\renewcommand\trace{\operatorname{Tr}}

\renewcommand{\d}{\mathrm{d}}

\usepackage{xcolor}

\newif\ifshowcorrections
\showcorrectionsfalse    

\newcommand{\correction}[1]{%
  \ifshowcorrections
    {\color{red}#1}%
  \else
    #1%
  \fi
}

\begin{document}
	
	\preprint{APS/123-QED}
	
	\title{Nonlinear squeezing of superpositions of quadrature eigenstates}
	
	\author{Vojt\v{e}ch Kucha\v{r}}
	\author{Petr Marek}%
	\affiliation{%
		Department of Optics, Palack\'y University, 17. listopadu 1192/12, 779 00 Olomouc, Czech Republic
	}%
	

	\date{\today}
	
	\begin{abstract}
		We introduce a family of operators exploiting the symmetry of superpositions of quadrature eigenstates (SQE) and demonstrate how the associated nonlinear squeezing, quantified by the expectation value of such operators, serves both as a witness of non-Gaussianity and as an indicator of the quality of SQE approximations. To establish the usefulness of this measure, we connect it to quantum state fidelity and evaluate its implications in breeding protocols. Finally, we construct optimal approximations of SQE states in truncated Fock spaces.
	\end{abstract}
	
	\maketitle
	
	
\section{Introduction}

Traveling light is a versatile experimental platform utilized for many applications of quantum information processing, such as quantum computation~\cite{Nielsen2012,asavanantOpticalQuantumComputers2022,bourassaBlueprintScalablePhotonic2021,larsenFaulttolerantContinuousvariableMeasurementbased2021b,guQuantumComputingContinuousvariable2009,ostergaardOctorailLatticeFourdimensional2025a,asavanantSwitchingfreeTimedomainOptical2023} quantum metrology~\cite{zhuangDistributedQuantumSensing2018,guoDistributedQuantumSensing2020,polinoPhotonicQuantumMetrology2020}, or quantum communication~\cite{braunsteinQuantumInformationContinuous2005, weedbrookGaussianQuantumInformation2012a, jouguetExperimentalDemonstrationLongdistance2013,hajomerLongdistanceContinuousvariableQuantum2024}. Since all of these applications are based on preparing and utilizing specific quantum states and their properties, preparation and characterization of quantum states, together with verification of their quantum features, is one of the important challenges in the field~\cite{parisQuantumStateEstimation2004,aolitaReliableQuantumCertification2015,chabaudCertificationNongaussianStates2021}. The states required by applications can be broadly classified into two categories. Gaussian squeezed light has fluctuations in a single quadrature reduced below the level of the vacuum state.  It can be directly applied in metrology~\cite{theligoscientificcollaborationGravitationalWaveObservatory2011,macconeSqueezingMetrologyUnified2020} or communication~\cite{furusawaUnconditionalQuantumTeleportation1998}, and is indispensable for manipulation of optical states~\cite{fiurasekGaussianTransformationsDistillation2002, yonezawaContinuousvariableQuantumInformation2010,hastrup2021b} and generation of large scale entangled states~\cite{menicucciUniversalQuantumComputation2006, yokoyamaUltralargescaleContinuousvariableCluster2013}. Despite its uses, squeezed light on its own is insufficient for the more advanced applications of quantum information technologies, such as quantum computation~\cite{nisetNogoTheoremGaussian2009}. For this, the Gaussian tool set of quantum optics needs to be supported by non-Gaussian quantum states serving as code states~\cite{gottesmanEncodingQubitOscillator2001,grimsmoQuantumComputingRotationsymmetric2020a,schlegelQuantumErrorCorrection2022,joshiQuantumInformationProcessing2021} or resources for operations~\cite{walsheContinuousvariableGateTeleportation2020,hastrupUnsuitabilityCubicPhase2021,konnoNoncliffordGateOptical2021}.

Superpositions of coherent states form one such class of useful non-Gaussian states with uses in quantum computation~\cite{guillaudRepetitionCatQubits2019,hastrupAllopticalCatcodeQuantum2022, regentHighperformanceRepetitionCat2023} and metrology~\cite{jooQuantumMetrologyEntangled2011,luisEquivalenceMacroscopicQuantum2001}. These states can be also considered for breeding of Gottesman-Kitaev-Preskill (GKP) states~\cite{Vasconcelos2010,weigandGeneratingGridStates2018,Hastrup2022}, but there is a catch --- the breeding protocol calls for nonphysical superpositions of quadrature eigenstates (SQE). Squeezing the superposed coherent states can approach this approximation, but for practical purposes, it is generally more efficient to directly aim at preparation of the already squeezed states. Both superpositions of coherent states and superpositions of displaced squeezed states were already experimentally prepared~\cite{ourjoumtsevGenerationOpticalSchrodinger2007,huangOpticalSynthesisLargeamplitude2015, etesseExperimentalGenerationSqueezed2015,sychevEnlargementOpticalSchrodingers2017,cotteExperimentalGenerationCoherentstate2022,
wangExperimentalPreparationManipulation2022,baoExperimentalPreparationGeneralized2022}, which opened up the problem of evaluating and characterizing quality of the states. Superposed coherent states can be evaluated by fidelity~\cite{Jozsa1994}. This is, however, impossible for approximations of quadrature eigenstates, due to the non-physical nature of infinitely squeezed states.  

Quality of the Gaussian quadrature squeezed light can be straightforwardly characterized by squeezing --- the value of the variance of the quadrature operator --- which often directly correlates with the noise added to the protocols~\cite{wangQuantumInformationGaussian2007}. Quantum non-Gaussian states, however, have properties given by complex features of their Wigner function, such as negative areas, or sub-Planck structure~\cite{zurekSubplanckStructurePhase2001, walschaersNongaussianQuantumStates2021}. In recent efforts, non-Gaussianity was also tied to directly measurable quantities which can be used to construct specific witnesses~\cite{miyataImplementationQuantumCubic2016}. A specific kind of such witness is nonlinear squeezing~\cite{brauerGeneralizedSqueezingWitness2025}, defined as the variance of a nonlinear function of quadrature operators~\cite{konnoNonlinearSqueezingMeasurementbased2021,kalaCubicNonlinearSqueezing2022}, which targets selected practically relevant non-Gaussian features~\cite{marekGeneralImplementationArbitrary2018,brauerGenerationQuantumStates2021, kalaGenuineContinuousQuantumness2025} and can replace fidelity when the ideal target state is nonphysical~\cite{marek2024}. Nonlinear squeezing provides valuable information and can be straightforwardly evaluated, but it has an up-front cost --- the design of a specific nonlinear operator.

In this paper we present a nonlinear squeezing operator aimed at evaluating the quality of approximated \correction{superpositions of quadrature eigenstates}. We construct optimal approximations of SQE states in truncated Fock spaces as ground states of this operator and connect the values of nonlinear squeezing to fidelity in interactions where SQE states act as a natural resource of non-Gaussianity. We perform numerical optimizations to obtain bounds for this fidelity and also for nonlinear GKP squeezing in breeding protocols, allowing us to make predictions about the outputs in relevant interactions solely based on SQE squeezing of input states.

\section{Superpositions of quadrature eigenstates}
Let $ \hat{a} $ be the bosonic annihilation operator of an electromagnetic field mode with canonically conjugate quadratures $ \hat{x}=\frac{1}{\sqrt{2}}\left(\hat{a}^\dagger+\hat{a}\right), $ $ \hat{p}=\frac{i}{\sqrt{2}}\left(\hat{a}^\dagger-\hat{a}\right), $ natural variables with $ \comm{\hat{x}}{\hat{p}}=i $ are used in this work.  With the exception of $\ket{x=0}, $ we use $ \hat{x}\ket{x}=x\ket{x}$ to denote eigenstates of quadrature operators. An ideal \correction{SQE} can be defined as

	\begin{equation}\label{eq:eigenket_superposition}
		\ket{C_{u,\infty,\varphi}}=\ket{u}+e^{i\varphi}\ket{-u},
	\end{equation}

where $ u\in\R, \varphi\in\left[0,2\pi\right). $ The cases $ \varphi = 0, \varphi = \pi $ will be referred to as even and odd superpositions respectively.\par These states appear as the asymptotic limit of superposed displaced squeezed states~\cite{Vasconcelos2010, Hastrup2022}, as the squeezing approaches infinity;

\begin{equation}\label{eq:squeezed_superposition}
\ket{C_{u,r,\varphi}}=\frac{1}{\sqrt{\mathcal{N}_{u,r}}}\left[\left(\hat{D}_x\left(u\right)+e^{i\varphi}\hat{D}^\dagger_x\left(u\right)\right)\hat{S}\left(r\right)\ket{0}\right],
\end{equation}

	where $ \ket{0} $ is the vacuum state, $ \hat{D}_x(u)=\exp(-iu \hat{p}) $  is the single-quadrature displacement operator, $ u\in\R, $ $ \hat{S}\left(r\right)=\exp\left[\frac{r}{2}\left(\hat{a}^2-\hat{a}^{\dagger 2}\right)\right] $ is the squeezing operator, $ r\in\R $, and $ \mathcal{N}_{u,r} $ is a normalization constant. SQE states, as defined in~\eqref{eq:eigenket_superposition}, are non-normalizable and thus unphysical, but their approximative physical versions~\eqref{eq:squeezed_superposition} are often considered as imperfect resources for quantum information processing protocols~\cite{gottesmanEncodingQubitOscillator2001,Vasconcelos2010,weigandGeneratingGridStates2018,Hastrup2022,schlegelQuantumErrorCorrection2022}. For this, it is important to be able to determine how well an arbitrary quantum state $ \hat{\rho}$ approximates the ideal state~\eqref{eq:eigenket_superposition}, which is relevant for theoretical analysis of quantum protocols~\cite{bourassaBlueprintScalablePhotonic2021,aghaeeradScalingNetworkingModular2025}, evaluation of experimentally prepared quantum states~\cite{endoNongaussianQuantumState2023,konnoLogicalStatesFaulttolerant2024}, and related optimization of quantum state preparation circuits. An effective way to evaluate quantum states is based on the ability to define the states as eigenstates of some hermitian operator~\cite{brauerGeneralizedSqueezingWitness2025}. The variance of the operator, often called squeezing, can then be used as a straightforward figure of merit and often directly quantifies the noise added in some quantum protocol.\par The general superpositions of quadrature eigenstatates~\eqref{eq:eigenket_superposition} can be defined as zero eigenvalue eigenstates of the operator
\begin{equation}\label{eq:o}
		\hat{O}\left(u,\varphi,c\right)=\hat{O}_x(u)+c\cdot\hat{O}_p(u,\varphi),
\end{equation}
with
	\begin{subequations}
		\label{eq:my_equations}
		\begin{align}
			\hat{O}_x(u) &= {\left(\hat{x}^2 - u^2\right)}^2 \label{eq:opx},\\
			\hat{O}_p(u,\varphi) &= \sum_{j=-\infty}^{+\infty} \ketbra{p=p_j}, \label{eq:opp}
		\end{align}
	\end{subequations}
where \(p_j = \frac{(2j-1)\pi-\varphi}{2u}\). Note that the operator $\hat{O}_p(u,\varphi)$ is not unique. For same cases, such as even and odd superpositions, it could be replaced by a modified parity operator~\cite{brauerCatabilityMetricEvaluating2025}. However, choosing the operator as~\eqref{eq:opp} allows us to evaluate the property of the state using only statistics of the $\hat{p}$ quadrature which can be easily accessed by homodyne detection. \correction{For the even case of the idealized approximation~\eqref{eq:squeezed_superposition} we find
\begin{equation}\label{}
	 \expval{\hat{O}(u,0,c)}=\frac{\frac{4e^{2r}u^2\left(e^{2r}u^2-3\right)}{e^{e^{2r}u^2}+1}+8e^{2r}u^2+3}{4e^{4r}},
\end{equation}
which is decreasing everywhere as a function of $ r $ and converges to zero as $ r\rightarrow+\infty, $ confirming that the ideal SQE is a ground state of this operator.}\par
Operator~\eqref{eq:o}, or rather a family of operators distinguished by different values of $c$, can be separated into parts involving only the $x$ and the $p$ quadrature representations, taking advantage of the symmetric nature of the target states~\eqref{eq:eigenket_superposition} and~\eqref{eq:squeezed_superposition}. The value of $c$ then serves as a balancing factor, quantifying the relative importance between the narrowness of the peaks visible in $x$ and the interference pattern visible in $p$. \correction{Note that \(\hat{O}_x\) is the square of a Hermitian operator and is, therefore, positive-semidefinite, while \(\hat{O}_p\) is positive-semidefinite due to being a sum of projectors. This implies that \(\hat{O}\left(u,\varphi,c\right)\) is positive-semidefinite as well.}\par
Similarly to the work done in~\cite{marek2024}, we interpret the mean value of operator~\eqref{eq:o} as nonlinear squeezing, \correction{because the operator is a positive-semidefinite observable and the target state is its ground state. Nonlinear squeezing \cite{brauerGeneralizedSqueezingWitness2025} can be generally defined as
\begin{equation}\label{eq:nlsq}
	\xi = \frac{\operatorname{var}_{\hat{\rho}_\mathrm{T}}[\hat{A}]}{\operatorname{var}_{\hat{\rho}_\mathrm{B}}[\hat{A}]},
\end{equation}
where \(\hat{\rho}_\mathrm{T}\) is some tested state, \(\hat{\rho}_\mathrm{B}\) is the benchmark state and \(\hat{A}\) is an observable constructed as a general non-linear function of \(\hat{x}\) and \(\hat{p}\). For the case of \(\hat{\rho}_\mathrm{B} = \ket{0}\) and \(\hat{A}=\hat{x}\), equation~\eqref{eq:nlsq} reduces to standard quadrature squeezing. In our case, keeping in mind that, for fixed mean values, the variances in~\eqref{eq:nlsq} are simply expectation values, we specifically define SQE squeezing as
\begin{equation}
	\xi_\mathrm{SQE} = \frac{\expval{\hat{O}(u,\varphi,c)}}{\min_{G}\expval{\hat{O}(u,\varphi,c)}},
\end{equation}
where \(\min_{G}\) denotes the minimum over the set of Gaussian states. Values of \(\xi_\mathrm{SQE}<1\) (or \(\xi_\mathrm{SQE}<0\text{ dB}\)) therefore imply non-Gaussianity of the tested state. This lower bound can be analytically expressed as
}
\correction{
\begin{widetext}
	\begin{align}\label{eq:gausslim}
		\min_{G}\expval{\hat{O}(u,\varphi,c)} = \min\left\{\min_{r\in\R}\left[\underset{E_A}{\underbrace{u^4 +\frac{3}{4} e^{-4r}-u^2e^{-2r}
		+\frac{uc}{\pi}~\vartheta_3\left(-\frac{\pi}{2},e^{-u^2 e^{2r}}\right)}}\right], \quad \underset{E_B}{\underbrace{\frac{uc}{\pi}}}\right\},
\end{align}
\end{widetext}}
where $ \vartheta_3\left(z,q\right)=\sum_{n=-\infty}^{+\infty}q^{n^2}e^{i2nz} $ is the third elliptic theta function. \correction{Here, \(E_A\) corresponds to \(\expval{\hat{O}\left(u,\varphi,c\right)}\) for a squeezed vacuum with finite squeezing \(r\in\R\) and \(E_B\) corresponds to the same expectation value for an infinitely squeezed displaced state at \(u\). These are the two possible Gaussian optima as confirmed by numerical optimization.} Nonlinear squeezing of SQE states therefore serves as a witness of non-Gaussian behavior.\par
Let us now compare the nonlinear squeezing for SQE states, defined as the expectation value of~\eqref{eq:o}, to other methods that can be used for quantifying quality of the quantum states. A common figure of merit often used for such analysis is the quantum state fidelity~\cite{Jozsa1994}
	\begin{equation}\label{eq:fidelity} F\left(\hat{\rho}_1,\hat{\rho}_2\right)={\left(\trace\sqrt{\sqrt{\hat{\rho}_1}\hat{\rho}_2\sqrt{\hat{\rho_1}}}\right)}^2,
	\end{equation}
where $ \hat{\rho}_1,\hat{\rho}_2 $ are density operators of two quantum states. However, using quantum state fidelity directly to evaluate the quality of approximative preparation of eigenket superposition approximations needs to be discouraged. The first reason is that the target state~\eqref{eq:eigenket_superposition} is not normalizable. Although~\eqref{eq:fidelity} can be calculated, it does not have any good physical meaning, as it is not from the interval $[0,1]$. The second reason is that, even if we consider the target state to be a specific finite approximation~\ref{eq:squeezed_superposition} with some value $r$, the issues do not disappear. This is because for any $r$ we could find $r'>r$ such that the fidelity of the two states is close to zero, even though the state with $r'$ better approximates the ideal state~\eqref{eq:eigenket_superposition}.\par
One way to avoid the problem of calculating fidelities with non-normalizable states, is to not look for the fidelity directly, but rather to consider utilizing the states~\eqref{eq:eigenket_superposition} and~\eqref{eq:squeezed_superposition} as interaction resources (virtually) and only then comparing the outputs, those being regular states with unit norm, using quantum state fidelity. The virtual operation that can be considered for this purpose, an operation for which the superposition of quadrature eigenstates is a natural resource, is the non-unitary operation
\begin{equation*}\label{}
    \hat{D}_x(u) + \hat{D}_x(-u).
\end{equation*}
This operation can be used to create the superposed coherent states, both regular and squeezed, but it can be also seen as a single step of the breeding protocol for GKP states~\cite{Vasconcelos2010}. This operation can be physically implemented only by Gaussian operations supplanted by the resource state~\eqref{eq:eigenket_superposition}. The full gate consists of a Gaussian coupling, either a beam splitter, or a quantum non-demolition (QND) interaction~\cite{Filip2005}, that couples the resource~\eqref{eq:eigenket_superposition} to the target state. This coupling is then followed by homodyne detection and post-selection on a specific outcome.\par
Formally, we can represent the full gate as follows. We will consider two modes of a bosonic field, with quadratures satisfying $ \comm{\hat{x}_j}{\hat{p}_k}=i\delta_{j,k}$, where $ j,k=1,2 $. The Gaussian unitary coupling will be represented by a unitary operator $\hat{U}_{\mathrm{INT}}$, which can either be the quantum nondemolition interaction~\cite{Filip2005}
	\begin{equation}\label{eq:qnd}
		\hat{U}_\mathrm{QND}=\exp(-i\hat{x}_1\hat{p}_2),
	\end{equation}
	or the balanced beam splitter~\cite{Dakna1997}
	\begin{equation}\label{eq:bs}
		\hat{U}_\mathrm{BS}=\exp\left[\frac{i\pi}{4}(\hat{p}_1\hat{x}_2-\hat{p}_2\hat{x}_1)\right].
	\end{equation}
	We shall specifically consider an interaction where the first mode contains the resource state $ \ket{\mathrm{R}}_1 $ and the second mode contains the vacuum state $ \ket{0}_2 $. The resource mode will then be measured by homodyne detection and the output state will be conditioned on measuring $ \hat{p}_1=0 $. The output state $ \ket{\Omega}_2 $ can then be, up to normalization, expressed as
\begin{equation}\label{}
    \ket{\Omega}_2 \propto\, _1\langle \hat{p}=0| \hat{U}_{\mathrm{INT}}|\mathrm{R},0\rangle_{12}.
\end{equation}

For the ideal resource state $ \ket{\mathrm{R}}=\ket{C_{u,\infty,\varphi}} $, we can straightforwardly calculate the evolution and arrive at
	\begin{equation}\label{eq:qndidealout} \ket{\Omega_{\mathrm{QND},u,\infty,\varphi}}=\left[\hat{D}_x\left(u\right)+e^{i\varphi}\hat{D}^\dagger_x\left(u\right)\right]\ket{0}
	\end{equation}
	for the QND interaction and
	\begin{equation}\label{eq:bsidealout} \ket{\Omega_{\mathrm{BS},u,\infty,\varphi}}=\left[\hat{D}_x\left(\frac{u}{\sqrt{2}}\right)+e^{i\varphi}\hat{D}^\dagger_x\left(\frac{u}{\sqrt{2}}\right)\right]\hat{S}\left(\frac{\ln 2}{2}\right)\ket{0}
	\end{equation}
for the balanced beam splitter, both up to normalization, where \(\ket{\Omega}\) is the state conditionally prepared in the remaining mode.\par
\begin{figure}[h]
	\includegraphics[width=\linewidth]{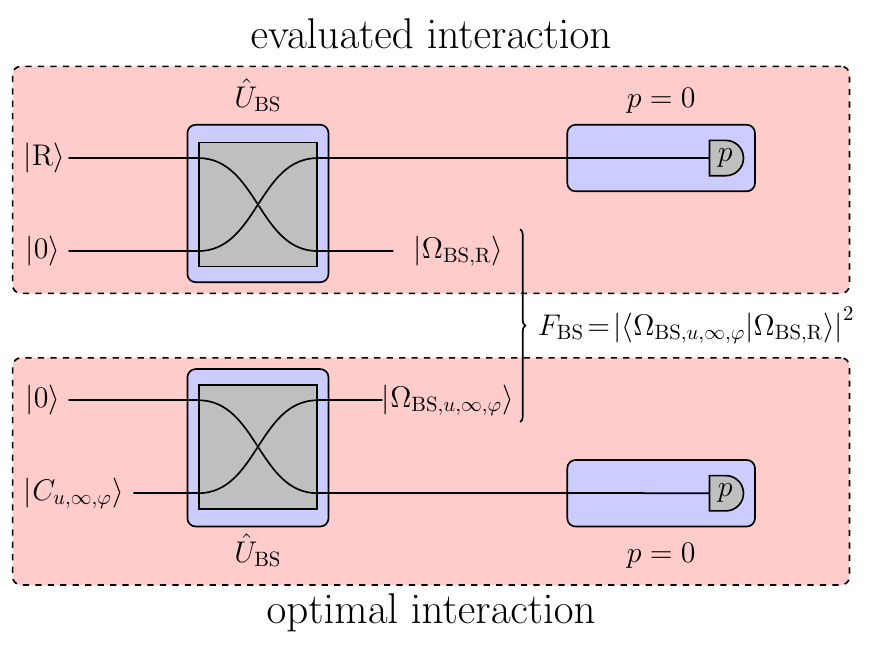}
	\caption{\label{fig:scheme} \emph{Schematic visualization of the virtual interaction-based method for evaluating SQE approximations.} Modes containing the evaluated state \(\ket{\mathrm{R}}\) and an ideal superposition of quadrature eigenstates \(\ket{C_{u,\infty,\varphi}}\) are split on identical beam splitters. Output states \(\ket{\Omega}\) are obtained by homodyne measurement of \(p=0\) in the remaining modes and subsequently utilized to calculate the virtual interaction based fidelity \(F_\mathrm{BS}\).}
\end{figure}

For any pure resource state $ \ket{\mathrm{R}} $  we can obtain the conditional output states in the $\hat{x}-$representation as
	\begin{align}\label{}
  &\ket{\Omega_{\mathrm{QND, R}}} = \frac{1}{\mathcal{N}_\mathrm{QND}} \iint\braket{x'}{\mathrm{R}}\braket{x-x'}{0}\ket{x} \d x \d x', \nonumber \\
&\ket{\Omega_{\mathrm{BS, R}}} = \frac{1}{\mathcal{N}_\mathrm{BS}}\iint \braket{\frac{x'-x}{\sqrt{2}}}{\mathrm{R}}\braket{\frac{x+x'}{\sqrt{2}}}{0}\ket{x} \d x \d x',
	\end{align}
where $\mathcal{N}_\mathrm{QND}$ and $\mathcal{N}_\mathrm{BS}$ are normalization coefficients. The process could be generalized for mixed states, but for our main purpose, benchmarking of the SQE squeezing, we can stick with pure states. The respective interaction fidelities can be now defined as
 \begin{align}\label{eq:Fidelities}
 &F_\mathrm{QND} = \abs{\braket{\Omega_{\mathrm{QND},u,\infty,\varphi}}{\Omega_{\mathrm{QND,R}}}}^2, \nonumber \\
 &F_\mathrm{BS} = \abs{\braket{\Omega_{\mathrm{BS},u,\infty,\varphi}}{\Omega_{\mathrm{BS,R}}}}^2,
 \end{align}
the calculation for \(F_\mathrm{BS}\) is visualized in Fig.~\ref{fig:scheme}. It can be shown that for the resource states that are superpositions of squeezed states $\ket{\mathrm{R}} = \ket{C_{u,r,\varphi}}$  (\ref{eq:squeezed_superposition}), the fidelities (\ref{eq:Fidelities}) are identical for both interactions and that in the limit as $r\rightarrow +\infty$ we get $F_\mathrm{QND} = F_\mathrm{BS} \rightarrow 1$. For simplicity, we shall from now on, consider a single virtual interaction fidelity $F = F_\mathrm{BS}$.

\section{Analysis}
If both \(F=F_\mathrm{BS}\) and \(\expval{\hat{O}(u,\varphi,c)}\) indicate how well a specific quantum state approximates the ideal SQE, it is reasonable to expect, that given the SQE squeezing value for a specific state, it should be possible to infer some information about the interaction fidelity. Specifically, obtaining a lower bound
\begin{equation}
	\begin{aligned}F_\mathrm{min}(z) &= \min_{\ket{\psi} \in S_z}F,\\S_z &= \left\{ \ket{\psi}\in\mathscr{H}_\infty : \langle\psi|\hat{O}(u,\varphi,c)|\psi\rangle = z\right\},
	\end{aligned}
\end{equation}
where the interaction fidelity is minimized over all pure states with \(\langle\hat{O}\rangle=z\), would allow us to predict how similarly the state in question behaves to the ideal SQE in the described virtual interactions based solely on its SQE squeezing.\par
Deriving the bound analytically is difficult due to the nature of the states that form the boundary, however, numerical methods can be used to obtain the bound with reasonable accuracy. The operator~\eqref{eq:o} is defined on an infinite dimensional Hilbert space \(\mathscr{H_\infty}\), but it can be truncated to a finite dimension~\cite{marek2024,provaznik2022}. Such finite dimensional representation can be used for evaluating the SQE squeezing on approximate states. Additionally, the ground states of such truncated operators posses the optimal SQE squeezing for any given dimension. Formally, the operator~\eqref{eq:o} truncated into a chosen finite dimensional truncated Fock space \(\mathscr{H}_N\), $N\in\N$ can be defined as
\begin{equation}\label{eq:truncated}
    \hat{O}^{[N]}(u,\varphi,c) =  \sum_{n=0}^{N-1}\ketbra{n} \hat{O}(u,\varphi,c) \sum_{n=0}^{N-1}\ketbra{n}.
\end{equation}
In the following, instead of utilizing the analytical expressions for \(\braket{x}{n}, \braket{p}{n}\) to calculate expectation values, we chose to approximate the constituents of $\hat{O}^{[N]}(u,\varphi,c)$; $ \hat{O}^{[N]}_x(u) $ and $ \hat{O}^{[N]}_p(a,\varphi) $, as matrices in the truncated Fock space. This leads to errors~\cite{provaznik2022}, however, it allows for much faster numerical optimization.\par
Implementing $ \hat{O}^{[N]}_x(u) $ is straightforward. Since this is an operator formed by fourth powers of $\hat{a}$ and $\hat{a}^{\dag}$, we can construct it with the help of the truncated annihilation operator $\hat{a}^{[N+4]}$ and then truncate it again.
Representing $ \hat{O}_p(u,\varphi) $ is more difficult, as quadrature eigenkets are not available in truncated Fock spaces. We can, however, utilize its $ \hat{p}- $representation, a series of delta distributions at $ p=\frac{(2j-1)\pi-\varphi}{2u}, j\in \Z$ to approximate it as
	\begin{align}\label{eq:opapprox}
		\tilde{\hat{O}}_p(u,\varphi)
		&= \frac{u}{\sqrt{\pi}}{(k)}_{\frac{1}{2}}{\left[\sin(u\hat{p}+\frac{\varphi}{2})\right]}^{2k}
	\end{align}
	where $ {(x)}_m=\frac{\Gamma(x+1)}{\Gamma(x-m+1)} $ is a generalization of the falling factorial and \(k\in\N\). 
	We choose \(k=100\) in the following numerical work, see Appendix~\ref{apendixa} for more details.\par
\begin{figure}
		\includegraphics[width=0.8\linewidth]{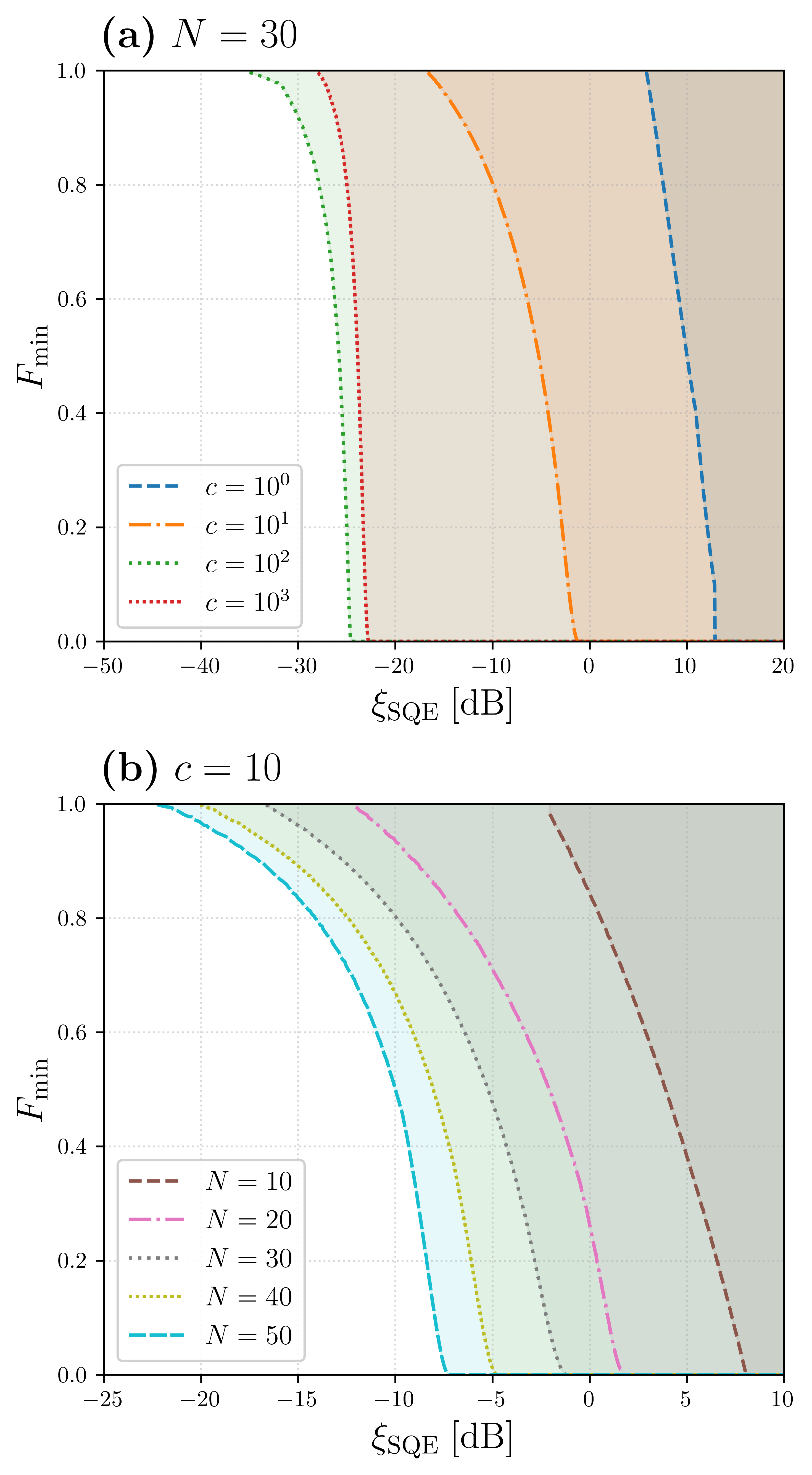}
		\caption{\label{fig:fbounds} \emph{Numerically optimized lower bounds for virtual interaction fidelity \(F_\mathrm{min}\) given SQE squeezing \(\xi_\mathrm{SQE}\).} Subplot \textbf{(a)} represents how the boundary changes with changes in relative weight \(c\) between the constituents of \(\hat{O}\left(u,\varphi,c\right)\), see \eqref{eq:o}. Subplot \textbf{(b)} represents how the boundaries with specific \(c\) differ with changes in dimensionality of the space \(N\).  The squeezing was evaluated in truncated Fock spaces using \(\hat{O}^{[N]}(3,0,c)\). The bounds represent worst-case scenarios, no states underneath the curves exist in their respective spaces. Therefore, we can infer minimum values of virtual interaction fidelity given SQE squeezing and vice versa.}
	\end{figure}

	Utilizing the finite dimensional approximations of \(\hat{O}\) in truncated Fock spaces, we can find the finite dimensional boundary
	\begin{equation}\label{eq:numbound}
		\begin{aligned}F_\mathrm{min}^{[N]}(z) &= \min_{\ket{\psi} \in S_z}F,\\S_z &= \left\{ \ket{\psi}\in\mathscr{H}_N : \langle\psi|\hat{O}\correction{^{[N]}}(u,\varphi,c)|\psi\rangle = z\right\},
		\end{aligned}
	\end{equation}
	as the Pareto frontier simultaneously minimizing both the virtual interaction fidelity and SQE squeezing. \correction{The Pareto frontier~\cite{jahanMulticriteriaDecisionAnalysis2016} is defined as the set of Pareto optimal points, i.e. all points, where virtual interaction fidelity cannot be lowered without increasing the expectation value \(z\). As such, it is equivalent to the lower bound defined in \eqref{eq:numbound}.} We have utilized the NSGA-II algorithm~\cite{Deb2002} to effectively find the frontier, our CUDA optimized Pymoo~\cite{Blank2020} implementation is available on GitHub~\cite{githubGitHubKucharonefopaff}. The SQE squeezing is expressed in decibels as
	\begin{equation}\label{eq:sqesq}
		\xi_\mathrm{SQE} = 10\log\left( \frac{\expval{\hat{O}^{[N]}(u,\varphi,c)}}{\min_{G}\expval{\hat{O}(u,\varphi,c)}} \right).
	\end{equation}
	The \(F_\mathrm{min}\) boundaries for various combinations of \(N, c\) are visualized in Fig.~\ref{fig:fbounds}. For each of the bounds, the filled area corresponds to the subspace where pure states can exist\correction{, and the dashed line represents the lower bound (Pareto frontier) itself}.\par
	SQE states are a crucial resource in breeding protocols for preparation of GKP state approximations~\cite{Vasconcelos2010,weigandGeneratingGridStates2018,Hastrup2022}. \correction{The ideal GKP qubit basis states are defined as~\cite{gottesmanEncodingQubitOscillator2001}
	\begin{equation}\label{eq:gkp}
		\ket{j_\mathrm{L}}=\sum_{k\in\Z}\ket{x=(2k-j)\sqrt{\pi}},\quad j=0,1.
	\end{equation}
	Although many similar encodings exist~\cite{joshiQuantumInformationProcessing2021}, in the GKP basis, all Clifford gates can be realized by Gaussian operations, making it especially suitable for optical quantum computers, with universal computation being achievable through magic state injection~\cite{garcia-alvarezBlochSpherePhaseSpace2021a}.} The \correction{approximations of \eqref{eq:gkp} obtained in breeding protocols} can, again, be evaluated by means of nonlinear squeezing~\cite{marek2024}
	\begin{equation}
		\xi_\mathrm{GKP_0} = 10\log\left( \frac{\expval{\hat{Q_0}^{[N]}}}{\min_{G}\expval{\hat{Q_0}}} \right),
	\end{equation}
	where
	\begin{equation}\label{key}
		\hat{Q}_0 = 2\sin^2\left(\frac{\hat{x}\sqrt{\pi}}{2}\right)+2\sin^2\left(\hat{p}\sqrt{\pi}\right).
	\end{equation}
	The expectation that states with lower SQE squeezing produce better GKP state approximations in breeding protocols can then be formally expressed as the existence of an upper bound
	\begin{equation}
		\begin{aligned}\xi_\mathrm{max}(z) &= \max_{\ket{\psi} \in S_z}\xi_{\mathrm{GKP_0}},\\S_z &= \left\{ \ket{\psi}\in\mathscr{H}_\infty : \langle\psi|\hat{O}(u,\varphi,c)|\psi\rangle = z\right\},
		\end{aligned}
	\end{equation}
	where \(\xi_\mathrm{GKP_0}\) is evaluated for a GKP state approximation obtained as the output of \(m\) rounds of a breeding protocol, where \(m^2\) identical copies of a state with SQE squeezing \(\xi_\mathrm{SQE}\) are used as the resource.\par
\begin{figure}
		\includegraphics[width=0.8\linewidth]{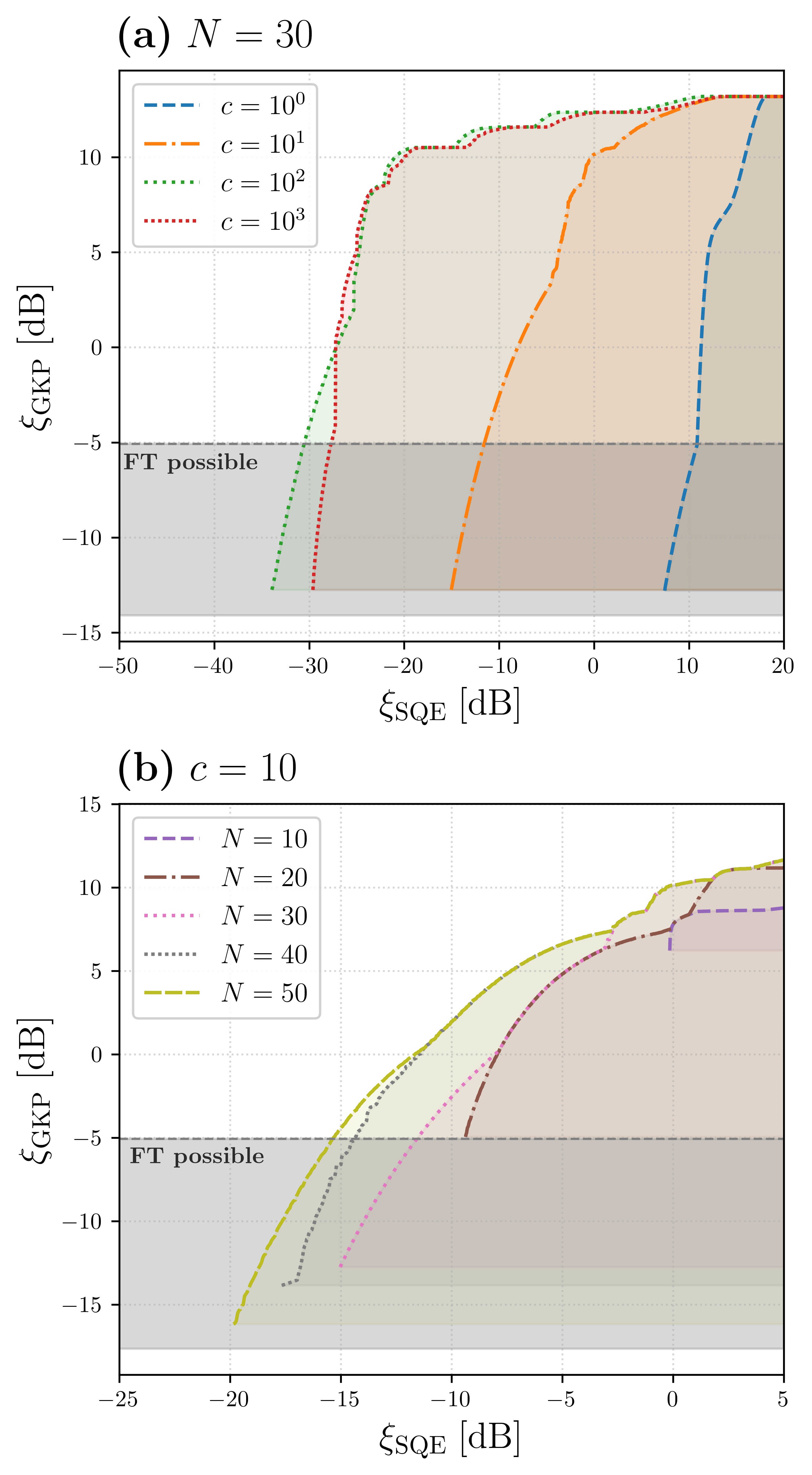}
		\caption{\label{fig:gkpbounds}\emph{Numerically optimized upper bounds for GKP squeezing after two rounds of breeding given SQE squeezing~\(\xi_\mathrm{SQE}\).} Subplot \textbf{(a)} represents how the boundary changes with changes in relative weight \(c\) between the constituents of \(\hat{O}\left(u,\varphi,c\right)\), see~\eqref{eq:o}. Subplot \textbf{(b)} represents how the boundaries with specific \(c\) differ with changes in dimensionality of the space \(N\). The squeezing was evaluated in truncated Fock spaces using \(\hat{O}^{[N]}(3,0,c)\). The area with gray filling represents the values of \(\xi_\mathrm{GKP_0}\) for which fault tolerance is possible~\cite{marek2024}. Therefore, the intersections of the boundaries with this area provide a necessary condition for \(\xi_\mathrm{SQE}\) for the resource states in order to breed GKP state approximations sufficient for fault tolerant quantum computation under ideal conditions.}
	\end{figure}

	Utilizing finite-dimensional approximations of both \(\xi_\mathrm{GKP_0}\) and \(\xi_\mathrm{SQE}\), the finite dimensional versions of this bound can be found in similar fashion to \(F_\mathrm{min}\) as the Pareto frontier of a multi-objective optimization problem with the objectives of simultaneously maximizing GKP squeezing and minimizing SQE squeezing. The boundaries are visualized in Fig.~\ref{fig:gkpbounds} for various combinations of \(N,c\). Each input state undergoes two rounds of the probabilistic breeding protocol introduced in~\cite{Vasconcelos2010} with postselection on \(p=0\). \correction{Similarily to Fig. \ref{fig:fbounds}, the shaded areas, each associated with one of the upper bounds (Pareto frontiers), correspond to subspaces where pure states possibly exist.}

	\section{Optimal SQE Approximations}
\begin{figure*}
		\includegraphics[width=\linewidth]{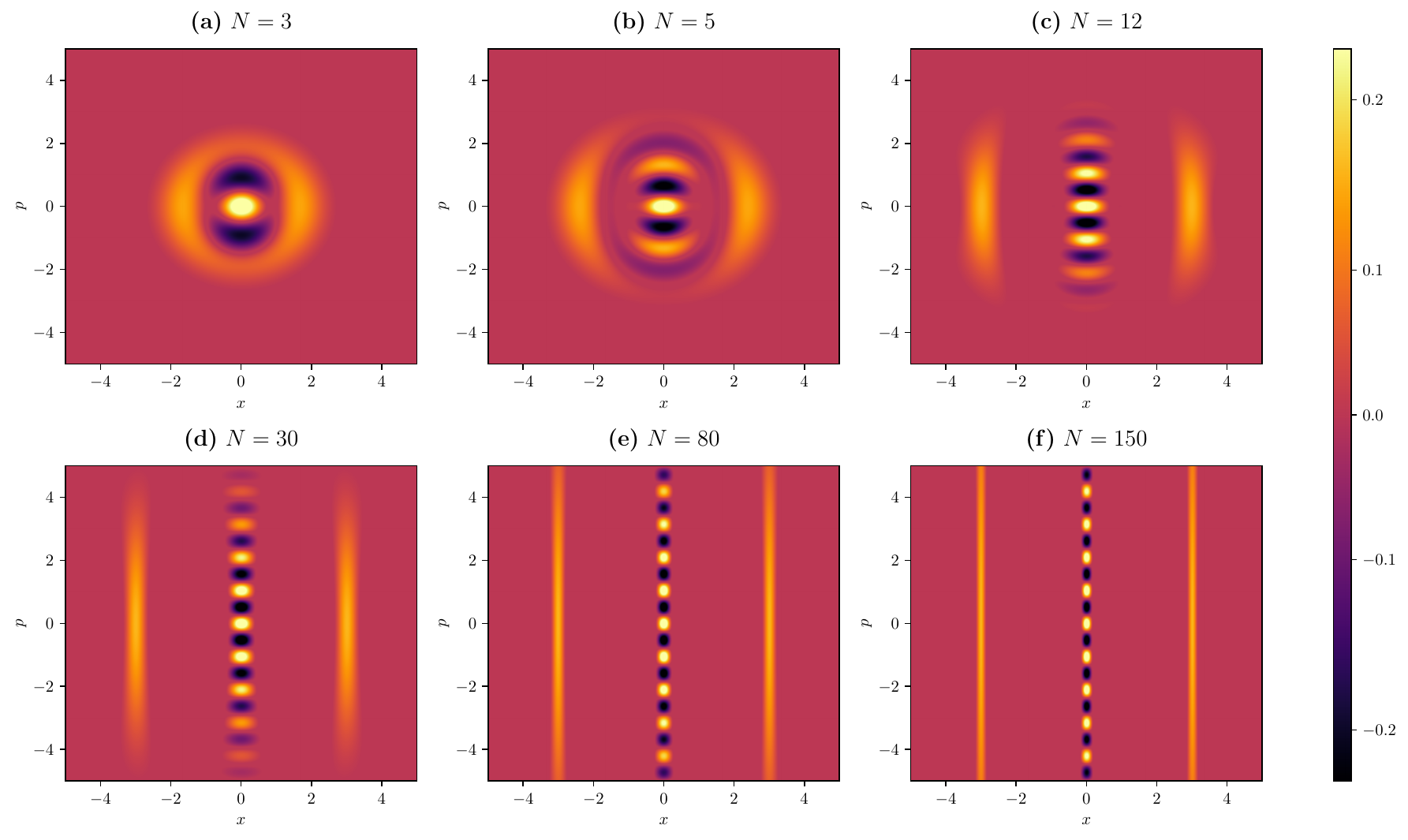}
		\caption{\label{fig:nstates} \emph{Optimal approximations for superpositions of quadrature eigenstates in truncated Fock spaces.} Each subplot visualizes the Wigner function of the state best approximating an even eigenket superposition $ \ket{x=3}+\ket{x=-3} $ in an $ N- $dimensional truncated Fock space. The plotted states were obtained as ground states of $ N- $dimensional representations of \(\hat{O}\left(u,\varphi,c\right)\) with \(u=3, \varphi = 0, c=10\)~\eqref{eq:o}. As such, these states minimize SQE squeezing~\eqref{eq:sqesq} in their respective truncated Fock spaces and are the optimal approximations of $ \ket{x=3}+\ket{x=-3} $ with stellar rank \correction{at most} \(r^\star\)~\eqref{eq:stellar}.} 
	\end{figure*}

	The analysis indicates that SQE squeezing can serve both as a witness of non-Gaussianity and as a measure of how well any state approximates a chosen superposition of quadrature eigenstates. We can utilize the measure-like properties of nonlinear squeezing to obtain optimal SQE approximations in truncated Fock spaces, which correspond simply to the ground states of~\eqref{eq:truncated} for any given dimension \(N\). Examples of such states \correction{for \(\varphi = 0\)} are plotted in Fig.~\ref{fig:nstates}.\par \correction{A common way to hierarchize non-Gaussian states is their stellar rank~\cite{chabaudStellarRepresentationNongaussian2020}, which is defined for a pure state \(\ket{\psi}\) as the number of zeros of its stellar function
	\begin{equation}
		F^\star_\psi = \exp\left( \frac{1}{2}\abs{\alpha}^2 \right)\braket{\alpha^*}{\psi},
	\end{equation}
	where \(\ket{\alpha}\) is the coherent state with amplitude \(\alpha\in\C\). It can be shown that the stellar rank of Fock states is equal to their photon number and that the stellar rank is invariant under Gaussian operations.}\par
	Given that the \correction{plotted} eigenstates are \correction{symmetric in the phase space, their Fock basis coefficients can be non-zero only for \(\ket{0},\ket{2},\dots\), and only up to the dimensional cut-off. This implies that} their stellar rank is \correction{at most}
	\begin{equation}\label{eq:stellar}
		r^\star = \begin{cases}
			N-2 &\text{for } N\bmod 2 = 0,\\
			N-1 &\text{otherwise.}
		\end{cases}
	\end{equation} Therefore, they represent optimal target approximations of superpositions of \correction{quadrature} eigenstates in quantum state preparation schemes with \correction{at most} \(r^\star\) single photon operations.\par\correction{In the context of state preparation, an alternate parametrization of \eqref{eq:o} can also be considered. For an ideal SQE \eqref{eq:eigenket_superposition}, Gaussian squeezing only affects the separation of the peaks, not their width. Choosing \(u = 1\) without loss of generality and subsequently transforming the operator by a squeezing operation we obtain 
\begin{equation}
		\hat{\mathcal{O}}\left(g,\varphi,c\right)=\hat{\mathcal{O}}_x(g)+c\cdot\hat{\mathcal{O}}_p(g,\varphi),
\end{equation}
with
	\begin{subequations}
		\begin{align}
			\hat{\mathcal{O}}_x(g) &= g^4{\left(\hat{x}^2 - \frac{1}{g^2}\right)}^2,\\
			\hat{\mathcal{O}}_p(g,\varphi) &= \sum_{j=-\infty}^{+\infty} \ketbra{p=\pi_j},
		\end{align}
	\end{subequations}
	where \(\pi_j = \frac{(2j-1)\pi-\varphi}{2/g}\) and \(g>0\) is the squeezing parameter. For any quantum state, we can now evaluate
	\begin{equation}
		\mathcal{O}(\varphi, c) = \min\limits_{g>0}\left[ \expval{\hat{\mathcal{O}}\left(g,\varphi,c\right)} \right]
	\end{equation}
	to find how well the state approximates an ideal SQE of any amplitude, if Gaussian squeezing is a free operation.\par
	This parametrization is more suitable when approximating SQEs using cat states, where both a larger separation of the peaks and stronger squeezing of individual peaks contribute to a better SQE approximation. 
	}

	\section{\label{results} Conclusion}
	We introduced a family of operators exploiting the properties of superpositions of quadrature eigenstates and established their usefulness within the framework of nonlinear squeezing. SQE squeezing serves both as a witness and measure of non-Gaussianity, and allows us to \correction{evaluate} how well any given state approximates a chosen symmetric superposition of quadrature eigenstates.\par
	Through numerical optimization, we discovered lower bounds for virtual interaction fidelity, quantifying how well a state with given SQE squeezing behaves as a resource for the non-unitary operation \(\hat{D}_x(u) + \hat{D}_x(-u)\) when compared to a superposition of quadrature eigenstates, the natural resource for such operation.\par
	In a similar fashion, we obtained upper bounds for GKP squeezing in a breeding protocol, which suggest that states with lower SQE squeezing generally produce better approximations of GKP states in such protocols.\par
	Furthermore, SQE squeezing allows us to construct optimal approximations for superpositions of quadrature eigenstates in truncated Fock spaces. These states represent optimal targets for experimental preparation with limited resources, effectively connecting theory with experiment and suggesting how well these intrinsically unphysical states can be approximated in the near future.

		\begin{acknowledgments}
We acknowledge support of the Czech Science Foundation (project 25-17472S). PM acknowledges  European Union's HORIZON Research and Innovation Actions under Grant Agreement no. 101080173 (CLUSTEC) and \correction{a grant from the Programme Johannes Amos Comenius under the Ministry of Education, Youth and Sports of the Czech Republic reg. no. CZ.02.01.01/00/22\_008/0004649.} VK acknowledges IGA-PrF-2025-010. We acknowledge use of the computational cluster at the Department of Optics. Data for generating the figures is openly available at \cite{kuchar_2025_15704165}.

		\end{acknowledgments}


\bibliography{references}		
\newpage
\onecolumngrid\appendix
\section{Truncated Fock space approximations\label{apendixa}}
In order to allow for effective optimizations, we approximate the constituents of \(\hat{O}(u,\varphi,c)\) as finite dimensional matrices expressed in the truncated Fock basis \(\left\{ \ket{0},\ket{1},\dots,\ket{N-1} \right\}\), where the truncated Fock space \(\mathscr{H}_N\), \(N\in\N\), is the linear hull of this basis. Given that both $ \hat{O}_x(u) $ and $ \hat{O}_p(a,\varphi) $ are defined in infinite dimensional spaces (see~\eqref{eq:opx} and~\eqref{eq:opp}), truncating these operators to finite dimensional matrices is generally associated with an error. The following is an analysis of the most dominant error in our approach.\par
To implement $ \hat{O}^{[N]}_x(u) $, we can directly utilize the truncated form of \(\hat{a}\) as outlined in the text. This leads to an excellent approximation with practically zero error in comparison to the error in $ \hat{O}^{[N]}_p(a,\varphi) $ and there is no reason to investigate further.\par
In the case of $ \hat{O}^{[N]}_p(a,\varphi) $, we are faced with the construction of quadrature eigenkets in a truncated Fock space. One could consider utilizing the expansion
\begin{equation}\label{ap:badeq}
	\ket{p_0} = \sum_{n=0}^{\infty}\ket{n}\!\!\braket{n}{p_0} = \sum_{n=0}^{\infty}\frac{{(-i)}^n e^{-{p_{0}}^2}}{\sqrt{2^n n! \sqrt{\pi}}}H_n(p_0)\ket{n},
\end{equation}
where \(H_n\) is the \(n-\)th Hermite polynomial, and truncating it to a finite sum capped at \(n=N\). This method unfortunately fails completely for some \(N\). While we did not investigate the cause thoroughly, our understanding of the eigenket \(\ket{p_0}\) as the infinite squeezing limit of a squeezed state with \(\expval{\hat{p}}=p_0\) points to the fact that while~\eqref{ap:badeq} converges to the eigenstate, its truncated form does not in any way guarantee \(\expval{\hat{p}}=p_0\). This leads to offsets of the individual projectors which cause the operator to behave in unintended ways.\par
To mitigate this issue, we have instead decided to approximate the infinite dimensional operator  $ \hat{O}_p(u,\varphi) $ using a different infinite dimensional operator  $ \tilde{\hat{O}}_p(u,\varphi)$, which can then be truncated in a similar manner to $ \hat{O}^{[N]}_x(u) $. Its form
\begin{equation}\label{ap:aprox}
	\tilde{\hat{O}}_p(u,\varphi) = \frac{u}{\sqrt{\pi}}{(k)}_{\frac{1}{2}}{\left[\sin(u\hat{p}+\frac{\varphi}{2})\right]}^{2k},
\end{equation}
is based on the fact that $ \hat{O}_p(u,\varphi) $ projects only onto $ p=\frac{(2j-1)\pi-\varphi}{2u}, j\in \Z$. The limit $\lim\limits_{k\rightarrow\infty}{\left[\sin(u\hat{p}+\frac{\varphi}{2})\right]}^{2k}$ is positive precisely at these points while being zero everywhere else. The multiplicative term ensures normalization around each of the peaks, i.e.
\begin{equation}\int_{\frac{j\pi-\varphi/2}{u}}^{\frac{j\pi-\varphi/2}{u}+\frac{\pi}{u}}\frac{u}{\sqrt{\pi}}{(k)}_{\frac{1}{2}}{\left[\sin(up+\frac{\varphi}{2})\right]}^{2k}\d p = 1, \quad\forall j\in\Z.
\end{equation}
Given the identical support and normalization, we can conclude that \(\lim\limits_{k\rightarrow\infty}\tilde{\hat{O}}_p(u,\varphi)=\hat{O}_p(u,\varphi)\). This limit acts as a justification for approximating \(\hat{O}_p(u,\varphi)\) in this manner, but does not find much use in practice, as numerical issues appear with large \(k\).\par
\begin{figure}[h]
	\includegraphics[width=\linewidth]{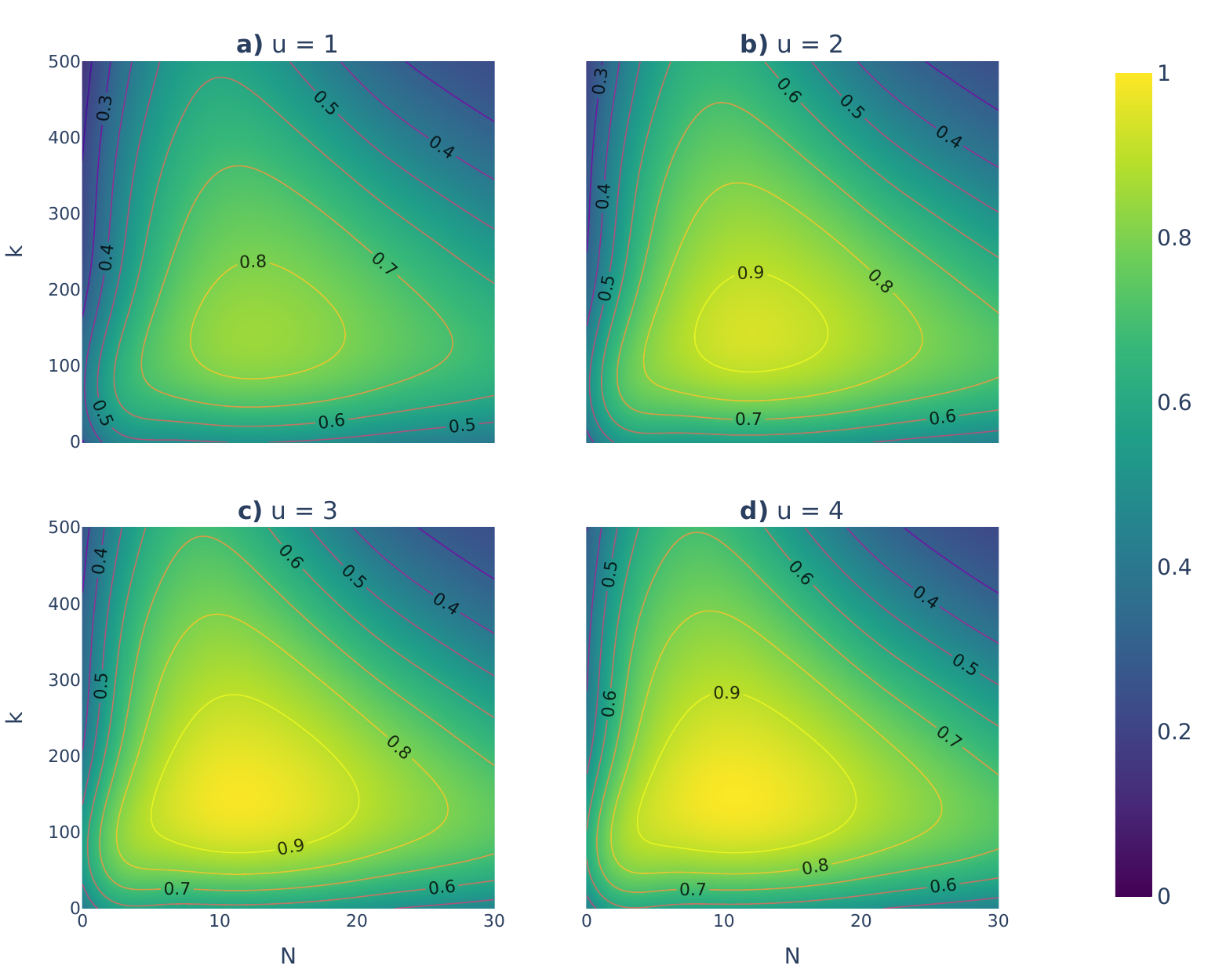}
	\caption{\label{fig:2derror} \emph{Accuracy of \(\tilde{\hat{O}}_p(u,\varphi)\) in infinite dimensional spaces.} Each plot represents the accuracy of the infinite dimensional approximation~\eqref{ap:aprox} for \textbf{a)} $u=1$, \textbf{b)} $u=2$, \textbf{c)} $u=3$, \textbf{d)} $u=4$, \(\varphi = 0\) in all cases. The approximation was calculated for individual Fock states and its accuracy is expressed as the normalized density of points, for which \(\abs{1-\expval{\tilde{\hat{O}}_p(u,0)}{n}/\expval{\hat{O}_p(u,0)}{n}}\leq 1\% \), i.e. if a single point is picked in the 0.9 contour, the probability of the approximation being off by more than 1 \% is smaller than 0.1.}
\end{figure}

In this case, the dominant error is the error associated with the approximation itself and not the truncation error, see Fig.~\ref{fig:2derror}. Given that the approximation numerically collapses for large \(k\), but also requires large \(k\) to work, we seek an optimal value that allows us to obtain reasonably accurate results for the optimizations we perform. Specifically, given the dimensionality of the truncated Fock spaces used and an analysis of the errors in Fig.~\ref{fig:2derror}, we settled for \(k=100\). Additionally, when looking for \(F_\mathrm{min}\) (see Fig.~\ref{fig:fbounds}), we utilized this data to choose \(u=3\).
\end{document}